\documentclass[preprint2]{aastex}

\def\kms{\ifmmode{\rm km\thinspace s^{-1}}\else km\thinspace s$^{-1}$\fi}

\slugcomment{To appear in Astrophysical Journal.}

\shorttitle{Absolute properties of CM~Dra}
\shortauthors{Morales et al.}

\begin{document}

\title{Absolute properties of the low-mass eclipsing binary CM~Draconis}

\author{Juan Carlos Morales\altaffilmark{1}, Ignasi 
Ribas\altaffilmark{1,2}, Carme Jordi\altaffilmark{1,3}, Guillermo 
Torres\altaffilmark{4}, Jos\'e Gallardo\altaffilmark{5}, Edward F. 
Guinan\altaffilmark{6}, David Charbonneau\altaffilmark{4}, Marek 
Wolf\altaffilmark{7}, David W. Latham\altaffilmark{4},
Guillem Anglada-Escud\'e\altaffilmark{8}, David H. Bradstreet\altaffilmark{9},
Mark E.\ Everett\altaffilmark{10}, Francis T. 
O'Donovan\altaffilmark{11}, Georgi Mandushev\altaffilmark{12} and
Robert D. Mathieu\altaffilmark{13}}

\date{Accepted for publication October 3rd, 2008}

\altaffiltext{1}{Institut d'Estudis Espacials de Catalunya (IEEC), Edif.\
Nexus, C/Gran Capit\`a 2-4, 08034 Barcelona, Spain; morales@ieec.uab.es}
\altaffiltext{2}{Institut de Ci\`encies de l'Espai (CSIC), Campus UAB, 
Facultat de Ci\`encies, Torre C5 - parell - 2a planta, 08193 Bellaterra, Spain}
\altaffiltext{3}{Departament d'Astronomia i Meteorologia, Institut de 
Ci\`encies del Cosmos, Universitat de Barcelona, C/ Mart\'i i Franqu\`es
 1, 08028 Barcelona, Spain}
\altaffiltext{4}{Harvard-Smithsonian Center for Astrophysics, 
60 Garden St., Cambridge, MA 02138, USA}
\altaffiltext{5}{Departamento de Astronom\'\i a, Universidad de Chile, 
Casilla 36-D, Santiago, Chile}
\altaffiltext{6}{Department of Astronomy and Astrophysics, 
Villanova University, PA 19085, USA}
\altaffiltext{7}{Astronomical Institute, Charles University, 
CZ-180 00 Praha 8, Czech Republic}
\altaffiltext{8}{Carnegie Institution of Washington, 
5241 Broad Branch Road NW, Washington DC, 20015, USA}
\altaffiltext{9}{Department of Astronomy and Physics, Eastern College, St. Davids, 
PA 19087, USA}
\altaffiltext{10}{Planetary Science Institute, 1700 E.\ Fort Lowell
Rd., Suite 106, Tucson, AZ 85719, USA}
\altaffiltext{11}{NASA Postdoctoral Program Fellow, 
Goddard Space Flight Center, 8800 Greenbelt Rd Code 690.3, Greenbelt,
MD 20771, USA}
\altaffiltext{12}{Lowell Observatory, 1400 West Mars Hill Road, Flagstaff, 
AZ 86001, USA}
\altaffiltext{13}{Department of Astronomy, University of Wisconsin-Madison,
475 North Charter Street, Madison, WI 53706, USA}

\begin{abstract} 

Spectroscopic and eclipsing binary systems offer the best means for
determining accurate physical properties of stars, including their
masses and radii. The data available for low-mass stars have yielded
firm evidence that stellar structure models predict smaller radii and
higher effective temperatures than observed, but the number of systems
with detailed analyses is still small. In this paper we present a
complete reanalysis of one of such eclipsing systems, CM~Dra, composed
of two dM4.5 stars. New and existing light curves as well as a radial
velocity curve are modeled to measure the physical properties of both
components. The masses and radii determined for the components of
CM~Dra are $M_{1}=0.2310\pm0.0009$~M$_{\odot}$,
$M_{2}=0.2141\pm0.0010$~M$_{\odot}$,
$R_{1}=0.2534\pm0.0019$~R$_{\odot}$, and
$R_{2}=0.2396\pm0.0015$~R$_{\odot}$. With relative uncertainties well
below the 1\% level, these values constitute the most accurate
properties to date for fully convective stars. This makes CM~Dra a
valuable benchmark for testing theoretical models. In comparing our
measurements with theory, we confirm the discrepancies reported
previously for other low-mass eclipsing binaries. These discrepancies
seem likely to be due to the effects of magnetic activity. We find
that the orbit of this system is slightly eccentric, and we have made
use of eclipse timings spanning three decades to infer the apsidal
motion and other related properties.

\end{abstract}


\section{Introduction}
\label{sec:introduction}

Late-type stars are the most common objects in the Galaxy, yet their
fundamental properties are still not well understood, in part because
their accurate measurement is challenging. Double-lined eclipsing
binary systems (hereafter EBs) have proven to be the best source of
accurate properties for low-mass stars, and a number of those systems
have already been studied in detail (see \citealp{Ribas2006} for a
review). These analyses have revealed that low-mass stars in EBs have
radii that are $\sim$10\% larger and effective temperatures that are
$\sim$5\% cooler than the predictions of stellar structure models. On
the other hand, their luminosities seem to agree well with model
calculations. These discrepancies have been attributed to the effects
of magnetic activity on the component stars
\citep[e.g.,][]{Torres2002,LopezMorales2005,Torres2006,LopezMorales2007,
Morales2008,Ribas2008}. Additional systems with accurately known stellar
properties that cover the entire range of sub-solar masses are needed
to better constrain the differences between models and observations.

CM~Draconis (hereafter CM~Dra, GJ~630.1A, $\alpha_{\rm
J2000.0}=16^{\rm h}34^{\rm m}20.\!^{\rm s}35$, $\delta_{\rm
J2000.0}=+57^{\circ}09'44.\!''7$) is a $V = 12.9$~mag EB system at a
distance of 14.5~pc from the Sun, which forms a common proper motion
pair with a $V=15$~mag white dwarf (GJ~630.1B, $\alpha_{\rm
J2000.0}=16^{\rm h}34^{\rm m}21.\!^{\rm s}57$, $\delta_{\rm
J2000.0}=+57^{\circ}10'09.\!''0$) at a separation of
$\sim$26~arcsec. This common proper motion pair moves at a relatively
large angular speed of roughly 2~arcsec per year, which may be
indicative of Population~II membership. Because of this, it has been
considered a useful system for estimating the primordial helium
abundance of the Universe through the detailed study of its components
\citep{Paczynski1984}.

CM~Dra was first investigated spectroscopically and photometrically by
\cite{Lacy1977}, and more recently by \cite{Metcalfe1996}. Both
studies indicate the system is composed of two similar dM4.5 stars with
masses of about 0.23 and 0.21 M$_{\odot}$, orbiting each other with a
period of 1.27~days.  \cite{Viti1997,Viti2002} estimated a metallicity
of $-1.0 < [M/H] < -0.6$ for the system, and inferred an effective
temperature of $3000 < T_{\rm eff} < 3200$~K. In this paper we
describe new observations of this binary that add significantly to the
body of existing measurements.  The unique position of CM~Dra as the
best known binary system composed of fully convective stars makes it
exceptionally important for testing models of such objects, and fully
justifies a reanalysis in the light of our new observations.

An important feature of this system is that, unlike other well-known
low-mass EBs, its orbit has a small but measurable eccentricity.  The
precise measurement of eclipse timings over a sufficiently long period
can thus potentially lead to the detection of apsidal motion. The rate
of this motion can be used to infer the internal structure constant
$k_{\rm 2}$ \citep{Kopal1978, Claret1993}, with which further tests of
the models are possible. In addition, the investigation of the times
of eclipse can also reveal the presence of third bodies in the system
through the light-time effect. Attempts to detect planets around
CM~Dra in this way have been carried out in the context of the Transit
of Extrasolar Planets Project (TEP, \citealt{Deeg1998, Deeg2000,
Doyle2000, Deeg2008}), although no compelling evidence of such objects
has been found as yet.

In this paper we present a thorough reanalysis of CM~Dra to determine
the fundamental properties of its components, including the masses,
radii and effective temperatures. Additionally, we have measured 
the rate of advance of the line of apsides, which turns out to be 
dominated by the General Relativity contribution. In the
following we describe first all available photometric and radial
velocity measurements. A combined analysis of all the information
using the Wilson-Devinney code \citep[hereafter WD,][]{WD1971} is
discussed in \S\,\ref{sec:lc_rv}, and in \S\,\ref{sec:tmin} the times
of minimum are used to estimate the apsidal motion. The
absolute properties of each component and the age and metallicity of
the system are derived in \S\,\ref{sec:properties}, and compared with
stellar model predictions in \S\,\ref{sec:models}. Finally, we
summarize our conclusions in \S\,\ref{sec:conclusions}.

\section{Time-series data for CM~Dra}
\label{sec:data_CMDra}
\subsection{Light curves}
\label{subsec:lc}

The photometric data available for CM~Dra come from a variety of
sources. In addition to making use of the original light curve in the
$I$ band published by \cite{Lacy1977}, six new light curves measured
in the $I$ and $R$ bands have been obtained with the 0.8m Four College
Automatic Photoelectric Telescope (hereafter FCAPT) located 
at Fairborn Observatory in southern Arizona in the Patagonia Mountains. 
Differential photoelectric photometry was conducted 
from 1995 - 2005 on 335 nights.  
The photometry was typically conducted using the Cousins R and I 
filters.  The primary comparison and check stars were HD 238580 and 
HD 238573, respectively. Integration times of 20-sec were used and 
the typical precision of the delta-R and -I band measures was 0.014~mag 
and 0.011~mag, respectively.  The relatively large uncertainties arise 
mainly from the faintness of the CM Dra and uncertainties in centering 
the variable star using blind-offsets (rather than direct acquires). 

\begin{table}[t]
\begin{center}
 \caption{Differential $R$- and $I$-band photometry for CM~Dra from
FCAPT. Dates are given in heliocentric Julian days on the TT time scale 
(HJED). The full version is available electronically.}
 \label{tab:FCAPT}
 \begin{tabular}{c c | c c}
 \tableline\tableline
 HJED       & $\Delta$R &  HJED        & $\Delta$I  \\ 
 \tableline                                         
2450172.99766 &  0.0243 &  2450172.99780 &  0.0651  \\
2450173.00000 &  0.0525 &  2450173.00015 &  0.0712  \\
2450173.00182 &  0.0297 &  2450173.00197 &  0.0604  \\
2450174.99836 &  0.0553 &  2450175.00271 &  0.0589  \\
2450175.00065 &  0.0711 &  2450175.96655 &  0.0513  \\
2450175.00257 &  0.0516 &  2450175.96882 &  0.0734  \\
2450175.96641 &  0.0036 &  2450175.97067 &  0.0569  \\
2450175.97052 &  0.0033 &  2450176.93331 &  0.0566  \\
2450176.93317 &  0.0230 &  2450176.93516 &  0.0612  \\
2450176.93502 &  0.0425 &  2450181.93217 &  0.0727  \\
  \tableline
 \end{tabular}
\end{center}
\end{table}

An additional light curve in the $r$-band was gathered with the Sleuth 
telescope located at the Palomar Observatory in southern California. 
Sleuth was one of three telescopes that together made up the 
Trans-atlantic Exoplanet Survey (TrES), and its primary use was to 
discover transiting planets orbiting stars brighter than $V=13$ 
\citep[e.g.,][]{ODonovan2006a, ODonovan2007,Mandushev2007}.  Sleuth 
consists of a lens with a physical aperture of 10~cm that images a field 
of view of size 5.7~degrees-square onto a thinned, back-illuminated CCD 
with 2048$\times$2048 pixels, corresponding to a plate scale of 
10~arcseconds per pixel.  From UT 2004 March 29 to UT 2004 June 6, Sleuth 
observed (as part of its survey for transiting planets) a field centered 
on the guide star HD~151613, and this field fortuitously contained our 
target CM~Dra.  Whenever weather permitted operation, the telescope 
gathered exposures in $r$-band with an exposure time of 90~s and a CCD 
readout time of 27~s, for a cadence of 117~s.  We used a photometric 
aperture of radius 30~arcseconds (3~pixels) to produce the differential 
photometric time series listed in Table \ref{tab:Sleuth} and shown in Fig. 
\ref{fig:fit_lc}. The calibration of TrES images, the extraction of the 
differential photometric time series (based on image subtraction methods), 
and the decorrelation of the resulting light curves is described elsewhere 
\citep{Dunham2004,Mandushev2005,ODonovan2006b}. The FCAPT and Sleuth data 
are collected in Table~\ref{tab:FCAPT} and Table~\ref{tab:Sleuth}.

\begin{table}[t]
\begin{center}
 \caption{Differential $r$-band photometry for CM~Dra from Sleuth. The
full version is available electronically.}
  \label{tab:Sleuth}
 \begin{tabular}{c c c}
 \tableline\tableline
 HJED          & $\Delta$r & $\sigma$    \\ 
 \tableline                                         
 2453093.80987 &  0.0194  &  0.0100 \\
 2453093.81120 &  0.0123  &  0.0080 \\
 2453093.81254 &  0.0013  &  0.0080 \\
 2453093.81387 & $-$0.0131  &  0.0070 \\
 2453093.81522 & $-$0.0041  &  0.0080 \\
 2453093.81656 & $-$0.0024  &  0.0070 \\
 2453093.81789 &  0.0106  &  0.0080 \\
 2453093.81923 &  0.0060  &  0.0080 \\
 2453093.82161 &  0.0058  &  0.0070 \\
 2453093.82294 & $-$0.0100  &  0.0070 \\
  \tableline
 \end{tabular}
\end{center}
\end{table}

\begin{figure}[t]
\centering
\includegraphics[width=\columnwidth]{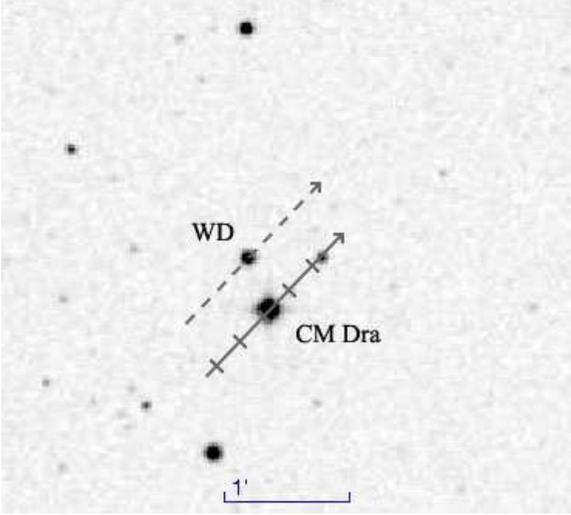}
\caption{POSS-II DSS2 image in the $R$ band showing the position of
CM~Dra at epoch 1991.5 and its proper motion on the sky. The common
proper motion white dwarf companion is labeled ``WD". Tick marks on
the path of CM~Dra are given in steps of 10 years from 1970 to
2010. North is up and East is left.}
\label{fig:track_CMDra}
\end{figure}

Collectively these light curves cover the observing seasons 1975,
1996--2001, and 2004, and add up to more than 20000 individual
measurements. The short orbital period of CM~Dra, along with its
possibly old age (see below), make it very likely that tidal
interactions have forced its components to rotate synchronously with
the orbital motion. Thus, although the stars are fully convective, it
is not surprising that they show a high level of chromospheric
activity as indicated, for instance, by the presence of flares
\citep[e.g.,][]{Eggen:67, Lacy:76, Nelson2007}.  Surface features
(spots) are also conspicuously present and are responsible for
modulations in the light curves that change from season to season.
This complicates the analysis significantly. Prior to combining the
different data sets, it is therefore necessary to correct the light
curves for these distortions. Additionally, the large proper motion of
CM~Dra on the sky is such that the system is approaching an
$R=16.5$~mag star to the NW, as shown in Fig.~\ref{fig:track_CMDra}.
Because different photometric apertures have been used to obtain the
measurements, the proximity of this star implies that the light curves
from different instruments may be affected to different degrees by
third light. This contamination must also be removed before the data
can be combined.

The correction for these spot effects and third light contribution was 
performed by carrying out preliminary fits to the light curves in each 
individual data set using the WD code. This program assumes a relatively 
simple spot model in which the features are circular and uniform. 
Nevertheless it is adequate as a first-order description. FCAPT $R$- and 
$I$-band data from the same season were used simultaneously. In the 
absence of spots, the model parameters in these fits that depend only on 
light curves are the eccentricity ($e$), the initial argument of the 
periastron ($\omega$), the inclination angle ($i$), the temperature ratio 
($T_{\rm eff,2}/T_{\rm eff,1}$, where subindex 2 indicates the less 
massive component), the surface pseudo-potentials ($\Omega_{i}$), which 
are related to the relative radii ($r_{i}$), and the passband-specific 
luminosity ratio ($\frac{L_{2}}{L_{1}}$). Properties that rely on the 
radial velocities, i.e., the semimajor axis ($a$), the mass ratio 
($M_{2}/M_{1}$) and the systemic radial velocity ($\gamma$), were held 
fixed at the values given by \cite{Metcalfe1996}. Limb darkening 
coefficients for these WD runs were computed for the standard Cousins $R$ 
and $I$ bands implemented in the code to account for possible corrections 
of these coefficients according to stellar properties at each iteration. 
The stellar atmosphere files in our WD implementation do not consider the 
Sloan $r$ band. Therefore, we carried out a number of tests to check the 
adequacy of assuming Cousins $R$ band for the Sleuth light curve. We did 
so by considering solutions incorporating the differential in the limb 
darkening coefficients between the Cousins $R$ band and the Sloan $r'$ 
band (compatible to Sloan $r$) calculated from \citet{Claret2004}.
No significant effects were 
found, owing to the fact that the light curve shape is quite insensitive 
to small variations in the darkening coefficients in the red bands.

The spot properties that can be solved for using the WD code are their
angular radius ($r_{\rm s}$), the longitude ($\phi_{\rm s}$), latitude
($\theta_{\rm s}$), and temperature contrast relative to the
photosphere ($T_{\rm s}/T_{\rm eff}$, where $T_{\rm s}$ represents the
spot temperature). Because $\theta_{\rm s}$, $r_{\rm s}$, and $T_{\rm
s}/T_{\rm eff}$ are strongly correlated and can usually not be
determined all at once, the procedure to fit the spots was carried out
by iterations. We first computed solutions with variable $\phi_{\rm
s}$ and $r_{\rm s}$ for several fixed values of $\theta_{\rm s}$,
trying both dark and bright spots with moderate temperature ratios.
Several scenarios with spots on one or both components were
tested. Upon reaching convergence, we fixed the values of $\phi_{\rm
s}$ and $r_{\rm s}$ and solved for $\theta_{\rm s}$ and $T_{\rm
s}/T_{\rm eff}$. In cases where convergence was not reached, we
selected the fits with fixed values of $\theta_{\rm s}$ and $T_{\rm
s}/T_{\rm eff}$ yielding the smallest residuals. In all of these fits
third light ($\ell_{3}$) was considered as a free parameter as well.
The solutions for the spot parameters and third light that give the
smallest residuals are shown in Table~\ref{tab:spots}. As seen, the
spot configurations change somewhat from season to season, providing
some evidence of either redistribution of the features or appearances
and disappearances. Third light is also seen to vary from data set to
data set for the reasons indicated above. In particular, the much
larger value for the Sleuth data reflects the large pixel scale of
that instrument, which makes contamination by neighboring stars more
likely.

\begin{table*}[t]
\begin{center}
 \caption{Spot and third light parameters from fits to the light
curves in each season. Third light is given as the percentage of the
total light coming from the system at phase 0.25. Parameters labeled
as fixed were obtained from the trial fits giving the best residuals.}
\label{tab:spots}
 \begin{tabular}{l c c c c c c c}
 \tableline\tableline
                  & \multicolumn{5}{c}{Spots}                                                                                    & \multicolumn{2}{c}{$\ell_{3}$ (\%)}                          \\
                  & Star      & $\theta$ ($^{\circ}$) & $\phi$ ($^{\circ}$) & $r_{\rm s}$ ($^{\circ}$) & $T_{\rm s}/T_{\rm eff}$ & R band & I band \\
 \tableline
 \citet{Lacy1977} 1975 & 1    & 21$\pm$8                &  76$\pm$5           & 42$\pm$3                 & 0.94$\pm$0.02    & --          & 1.3$\pm$0.8 \\
 FCAPT 1996       & 2         & 45 (fixed)              & 338$\pm$6           & 13$\pm$1                 & 1.09 (fixed)     & 4.1$\pm$1.2 & 2.3$\pm$1.2 \\
 FCAPT 1997       & 1         & 30 (fixed)              & 316$\pm$7           & 32$\pm$6                 & 0.96 (fixed)     & 3.4$\pm$1.2 & 3.0$\pm$1.2 \\
                  & 2         & 30 (fixed)              & 304$\pm$12          & 12$\pm$5                 & 1.09 (fixed)     &             &             \\
 FCAPT 1998       & 1         & 30 (fixed)              & 315$\pm$7           & 40$\pm$2                 & 0.96 (fixed)     & 4.4$\pm$0.8 & 3.3$\pm$0.8 \\
 FCAPT 1999       & 1         & 45 (fixed)              & 119$\pm$11          & 15$\pm$3                 & 1.09 (fixed)     & 4.8 (fixed) & 2.9 (fixed) \\
                  & 1         & 45 (fixed)              & 255$\pm$11          & 19$\pm$7                 & 0.96 (fixed)     &             &             \\
 FCAPT 2000       & \multicolumn{5}{c}{Spot modulation not significant}                                                     & 5.5$\pm$1.7 & 3.6$\pm$1.7 \\
 FCAPT 2001       & 1         & 30 (fixed)              & 297$\pm$8           & 23$\pm$3                 & 1.09 (fixed)     & 1.4$\pm$1.7 & 1.6$\pm$1.7 \\
 Sleuth 2003      & 2         & 45 (fixed)              & 273$\pm$2           & 32$\pm$1                 & 0.96 (fixed)     &12.3$\pm$0.9 & --          \\
  \tableline
 \end{tabular}
\end{center}
\end{table*}

\begin{figure}[t]
\centering
\includegraphics[width=\columnwidth]{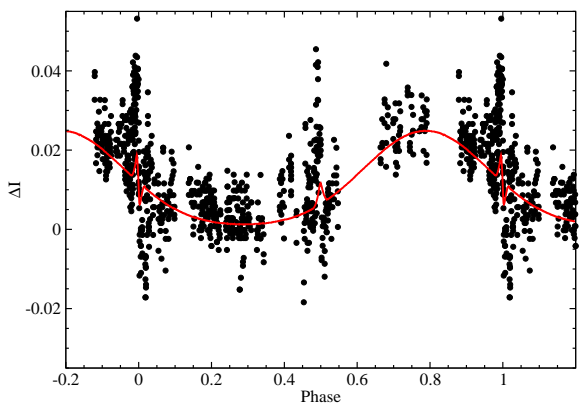}
\caption{Differential effect of star spots on the $I$-band light curve
of \citet{Lacy1977}. The solid line represents the model described in
the text.}
\label{fig:spot_mod}
\end{figure}

The corrections for spots and third light in each season were
computed from the difference between the theoretical curves from
these full fits and synthetic curves calculated with the same
geometric and radiative parameters but with no spots and no third
light. We then subtracted these effects from the original data. As an
example, Fig.~\ref{fig:spot_mod} shows the differential effect of the
spots for the light curve of \citet{Lacy1977}.

With these transitory effects removed, the photometric data can be
combined more easily for analysis with WD. For practical reasons, we
found it convenient to bin the large number of original data points in
order to reduce the computing time for the light curve solutions. The
relevant information resides almost completely in the eclipse phases
(it depends mostly on their detailed shape) so that averaging outside
of eclipse has essentially no impact on the results. We therefore
averaged the observations outside of eclipse from the same instruments
into bins of 0.04 in phase. This procedure was applied to the FCAPT
observations and the Sleuth observations. The total number of points
used in the solutions is 5356. Unit weight was assigned to observations
that have no reported errors, as is the case for the FCAPT data and
also \citet{Lacy1977}, whereas individual weights were used for the
Sleuth observations, for which internal errors are available. For the
out-of-eclipse averages from FCAPT and Sleuth we adopted as weights
the number of combined points and the reciprocal of the standard
deviation squared, respectively.

\subsection{Radial velocity data}
\label{subsec:vc}

For the present study we have made use of the same spectroscopic material 
discussed by \cite{Metcalfe1996}, obtained over a period of nearly 5 years 
with an echelle spectrograph on the 1.5m Tillinghast reflector at the F.\ 
L.\ Whipple Observatory (Mount Hopkins, Arizona). These observations were 
taken at a resolving power $\lambda/\Delta\lambda \approx 35,\!000$, and 
cover approximately 45\,\AA\ in a single order centered near the Mg~I $b$ 
triplet at $\sim$5187\,\AA.  For further details we refer the reader to 
the work of \cite{Metcalfe1996}. Here we have reanalyzed these spectra 
with improved techniques compared to the original study.  Radial 
velocities were obtained with TODCOR \citep{Zucker1994}, a two-dimensional 
cross-correlation algorithm. The template for both components was chosen 
to be an observation of Barnard's star (GJ~699, M4Ve) taken with a similar 
instrumental setup, which provides a close match to the spectral type of 
CM~Dra. Unlike the original study, here we have made a special effort to match 
the rotational broadening of each component by convolving the spectrum of 
Barnard's star (assumed to have negligible rotation) with a standard 
rotational profile. The values of the projected rotational velocity of the 
components ($v \sin i$) that provide the best match to the stars are $9.5 
\pm 1.0$~\kms\ for the primary and $10.0 \pm 1.0$~\kms\ for the secondary. 
The average light ratio derived from these spectra is $L_{2}/L_{1} = 0.91 
\pm 0.05$ at the mean wavelength of our observations.

As a test, we experimented with other templates obtained with the same
instrumentation to investigate the possibility of systematic errors in
the velocities due to ``template mismatch'' \citep[see,
e.g.,][]{Griffin2000}, which might bias the mass determinations. The
use of a template made from an observation of the star GJ~725\,A
(M3.5V) produced rather similar velocities, and an orbital solution
with nearly identical elements and formal uncertainties only slightly
higher than our previous fit.  The minimum masses from this solution
were smaller than our previous results by only 0.23\% and 0.14\% for
the primary and secondary, respectively, which are below the formal
errors in those quantities. A template from an observation of GJ~51
(M5.0V) gave an orbital solution that was significantly worse, and
minimum masses 0.67\% and 0.72\% higher than those from our reference
fit. As a measure of the closeness of the match to the real components
of CM~Dra, we computed for each template the cross-correlation value
from TODCOR averaged over all exposures.  Both of the alternate
templates, which bracket the spectral type of CM~Dra, gave average
correlation values that were lower than we obtained with the GJ~699
template (particularly for GJ~51), indicating the match is not as
good. The results using Barnard's star are thus preferable, and the
above tests indicate template mismatch is unlikely to be significant.

The spotted nature of the CM~Dra implies the possibility of systematic 
effects on the measured radial velocities that could bias the inferred 
masses and radii of the stars. In principle the WD code can approximately 
take into account these effects in solutions that use spectroscopic and 
photometric observations simultaneously, as long as those observations are 
contemporaneous.  Unfortunately, this is not the case here, and as seen in 
Table~\ref{tab:spots} the properties of the spots change significantly 
with time.  In order to at least provide an estimate of the effect, we 
have performed experiments in which we perturbed the individual velocities 
by adding the radial-velocity corrections that WD computes for each of the 
spot configurations in Table~\ref{tab:spots}. We then carried out 
Keplerian fits in each case, and we compared them. The differences in the 
key parameters (i.e., the minimum masses $M \sin^3 i$, projected semimajor 
axis $a \sin i$, $e$, $\omega$, and $M_{2}/M_{1}$) are always within the 
formal errors. This is not surprising, given that the individual velocity 
corrections are typically smaller than 0.2~\kms. Nevertheless, to be 
conservative, we have taken half of the maximum difference in each 
parameter as a measure of the possible systematic effect due to spots, 
and added this contribution in quadrature to the uncertainties determined 
from the analysis described in \S\,\ref{sec:lc_rv}.

The measured radial velocities in the heliocentric frame are listed in 
Table~\ref{tab:rv}, without any corrections. They supersede the 
measurements reported by \cite{Metcalfe1996}. The median uncertainties are 
approximately 1.2~\kms\ and 1.4~\kms\ for the primary and secondary, 
respectively. \cite{Metcalfe1996} did not report individual errors for
their radial velocities, but we may take the rms residuals from their
orbit as representative values. Compared to those (1.77~\kms\ and
2.33~\kms\ for the primary and secondary, respectively), our velocities
give significantly smaller residuals (1.30~\kms\ and 1.40~\kms;
Table~\ref{tab:fit_rv}). We attribute this to our use of templates
that better match the rotational broadening of each component (see
above), whereas \cite{Metcalfe1996} used an unbroadened template.

\begin{table*}[!ht]
\begin{center}
 \caption{Radial velocity measurements for CM~Dra in the heliocentric
frame. The full version of this table is available electronically.}
\label{tab:rv}
 \begin{tabular}{c c c c c}
 \tableline\tableline
 HJED      & $v_{rad,1}$ (\kms) & $\sigma_{1}$ (\kms) & $v_{rad,2}$ (\kms) & $\sigma_{2}$ (\kms)\\ 
 \tableline
 
2445158.7745 &   -74.57 & 1.20 &  -164.13 & 0.96 \\
2445783.8997 &  -140.28 & 0.12 &   -97.84 & 0.10 \\
2445783.9023 &  -136.97 & 0.43 &   -98.00 & 0.34 \\
2445783.9033 &  -134.58 & 0.12 &   -99.67 & 0.10 \\
2445783.9068 &  -134.07 & 0.12 &   -97.46 & 0.10 \\
2445783.9110 &  -131.36 & 0.91 &  -101.19 & 0.72 \\
2445783.9187 &  -128.69 & 0.49 &  -104.71 & 0.39 \\
2445783.9314 &  -125.45 & 0.49 &  -107.25 & 0.39 \\
2445783.9457 &  -125.46 & 0.49 &  -115.88 & 0.39 \\
2445783.9690 &  -113.92 & 0.49 &  -122.53 & 0.39 \\
  \tableline
 \end{tabular}
\end{center}
\end{table*}

\section{Analysis of light and radial velocity curves}
\label{sec:lc_rv}

Prior to combining these curves, the times of observation were
transformed to the uniform Terrestrial Time (TT) scale in order to
avoid discontinuities resulting from the more than 30 leap seconds
that have been introduced in the interval spanned by the various
data sets. For the analysis in this section we used the
2005 version of the 2003 WD code although updates in this version
do not affect the fitting mode used for detached binaries.
The program models proximity effects in
detail, although they are negligible for a well-detached system such
as CM~Dra. The reflection albedos for both components were held fixed
at the value 0.5, appropriate for convective envelopes, and a gravity
brightening coefficient of 0.2 was adopted following
\citet{Claret2000}. For the limb darkening we adopted the square root
law, with coefficients computed dynamically at each iteration from the
{\sc phoenix} atmosphere models \citep{Allard1995}, in order to follow
the evolution of the components' properties.

The light and velocity curves were adjusted simultaneously with WD
solving for the epoch of primary eclipse ($T_{0}$), the eccentricity
($e$), the argument of the periastron ($\omega$), the inclination
($i$), the semimajor axis ($a$), the systemic radial velocity
($\gamma$), the mass ratio ($M_{2}/M_{1}$), the secondary effective
temperature ($T_{\rm eff,2}$), the luminosity ratio at each bandpass
($L_{2}/L_{1}$), and the surface potentials ($\Omega_{i}$). To first
order the light curves are only sensitive to the temperature
\emph{ratio} of the components. Because the limb darkening
coefficients need to be interpolated from theoretical tables, we
assumed $T_{\rm eff,1}=3100$~K according to the results of
\citet{Viti1997, Viti2002} and fitted for the value of $T_{\rm
eff,2}$.

Given that the data span over 30 years, we initially attempted also to
estimate the period ($P$) as well as the apsidal motion rate
($\dot{\omega}$) with WD directly from the light curves,
simultaneously with the other adjustable quantities. We found that
this did not yield satisfactory results, and the value for
$\dot{\omega}$ was statistically insignificant compared to its large
error. We then chose to set $\dot{\omega}$ to zero and fit each of the
light curves separately in order to minimize the effects of possible
changes in $\omega$ from epoch to epoch. The period was held fixed at
the value found in the analysis of eclipse timings described later in
\S\,\ref{sec:tmin}, which is $P = 1.268389985\pm0.000000005$ days
(this value \emph{does} account for the small effect of
$\dot{\omega}$, as described below). We solved for the parameters of
each light curve, then computed the weighted averages, and
subsequently solved for the parameters of the radial velocity
curves. This was iterated until convergence, as judged by the changes
from one iteration to the next compared to the internal errors
reported by the WD code, i.e., convergence is reached when
corrections are smaller than errors.

\begin{table*}[!ht]
\begin{center}
\footnotesize
 \caption{Light-curve solutions for CM~Dra from the different data sets. The period adopted is $P=1.268389985$~days.}
 \label{tab:fit_lc}
 \begin{tabular}{l c c c c}
 \tableline\tableline
 Parameter                                           & Lacy         & FCAPT         & Sleuth        & Average        \\
\tableline
 \multicolumn{5}{l}{Physical properties}                              \\
  \hspace{0.1cm} $T_{\rm 0}$ (HJD$-$2400000)         & 42958.620510(24)   & 51134.661970(13)  & 53127.302690(21)   & 48042.32743(24)\tablenotemark{a}\\
  \hspace{0.1cm} $e$                                 & 0.00521(56)        & 0.00686(50)       & 0.00424(56)        & 0.0054(13)              \\
  \hspace{0.1cm} $\omega_{\rm 0}$ ($^{\circ}$)       & 108.1(2.2)         & 101.9(0.9)         & 113.9(3.8)         & 107.6(6.3)                  \\
  \hspace{0.1cm} $i$ ($^{\circ}$)                    & 89.784(64)        & 89.770(28)        & 89.712(62)        & 89.769(73)               \\
  \hspace{0.1cm} $\Omega_{1}$                        & 15.736(50)         & 15.877(39)         & 15.862(61)         & 15.79(11)                 \\
  \hspace{0.1cm} $\Omega_{2}$                        & 15.631(59)         & 15.506(40)         & 15.582(75)         & 15.59(10)                 \\
  \hspace{0.1cm} $r_{1}$\tablenotemark{b}            & 0.06757(12)        & 0.06700(12)        & 0.06690(17)        & 0.0673(5)              \\
  \hspace{0.1cm} $r_{2}$\tablenotemark{b}            & 0.06350(17)        & 0.06403(12)        & 0.06377(17)        & 0.0637(4)              \\
\\
\multicolumn{5}{l}{Radiative properties}                              \\
  \hspace{0.1cm} $T_{\rm eff,\it 1}$ (K)                & \multicolumn{3}{c}{3100 (fixed)}              &   \\
  \hspace{0.1cm} $T_{\rm eff,\it 2}/T_{\rm eff,\it 1}$ & 0.9984(7)        & 0.9926(4)        & 0.9923(5)        & 0.9960(40)            \\
  \hspace{0.1cm} $\left( L_{2}/L_{1} \right)_{\rm R}$ (ph. 0.25)       & --            & 0.8721(32)        & 0.8632(63)        & 0.8654(89)            \\
  \hspace{0.1cm} $\left( L_{2}/L_{1} \right)_{\rm I}$ (ph. 0.25)       & 0.8764(43)        & 0.8782(33)        & --            & 0.8768(44)            \\
  \hspace{0.1cm} Albedo                                & \multicolumn{3}{c}{0.5 (fixed)}               &   \\
  \hspace{0.1cm} Gravity darkening                     & \multicolumn{3}{c}{0.2 (fixed)}               &   \\
\\
\multicolumn{5}{l}{Limb darkening coefficients (square root law)}                    \\
  \hspace{0.1cm} $x_{1}$ \& $y_{1}$ $R$            & \multicolumn{3}{c}{0.268 \& 0.690}    &  \\
  \hspace{0.1cm} $x_{2}$ \& $y_{2}$ $R$            & \multicolumn{3}{c}{0.293 \& 0.669}    &  \\
  \hspace{0.1cm} $x_{1}$ \& $y_{1}$ $I$            & \multicolumn{3}{c}{$-$0.043 \& 1.011} &  \\
  \hspace{0.1cm} $x_{2}$ \& $y_{2}$ $I$            & \multicolumn{3}{c}{$-$0.018 \& 0.991} &  \\
\\
\multicolumn{5}{l}{Other quantities pertaining to the fits}                       \\
  \hspace{0.1cm} $\sigma_R$ (mag)                         & --                 & 0.0236       & 0.0137       &      \\
  \hspace{0.1cm} $\sigma_I$ (mag)                         & 0.0071             & 0.0130       & --            &       \\
  \hspace{0.1cm} $N_{\rm obs}$                            & 830          & 1656 (R) , 1691 (I) & 1179          &                \\
 \tableline
 \end{tabular}
\tablenotetext{a}{Reference epoch of each light curve corrected to a central epoch.}
\tablenotetext{b}{Volume radii.}
\end{center}
\end{table*}

\begin{figure*}[t]
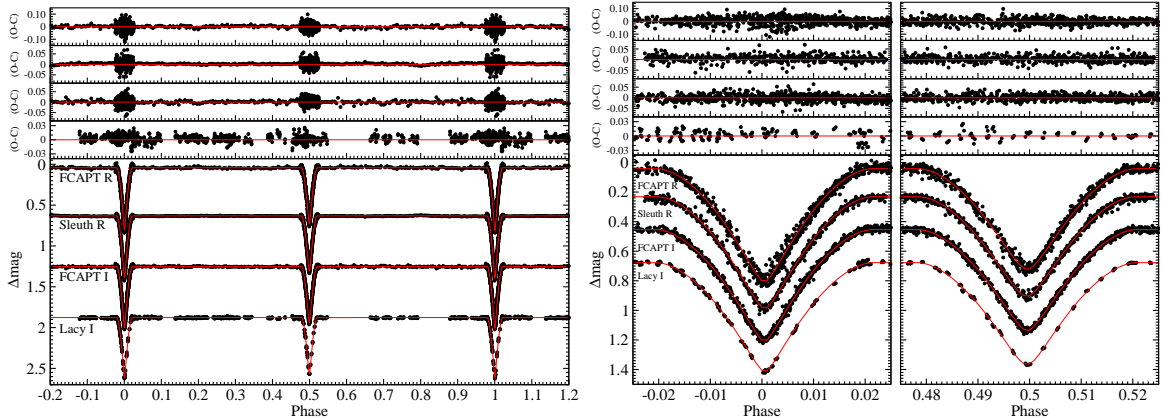

\centering
\includegraphics[width=\columnwidth]{f3a.eps}
\includegraphics[width=\columnwidth]{f3b.eps}
\caption{Left: Rectified light curves of CM~Dra after subtracting the
effects of third light and spots, separated by instrument.
Observations outside of eclipse are binned as described in the
text. Residuals are shown at the top.  Right: Enlargement around the
eclipse phases. All light curves are plotted as differential magnitude
vs.\ phase, and residuals are plotted in the same order as the light
curves. Note the different scales for the residuals of each
instrument.}
\label{fig:fit_lc}
\end{figure*}

\begin{table}[t]
\begin{center}
 \caption{Spectroscopic solution for CM~Dra. Period is held fixed at the value $P=1.268389985$~days.}
 \label{tab:fit_rv}
 \begin{tabular}{l c}
 \tableline\tableline
 Parameter                              & Value                       \\
\tableline
 \multicolumn{2}{l}{Physical properties}                              \\
  \hspace{0.1cm} $T_{\rm 0}$ (HJD)                     & 2446058.56471$\pm$0.00026   \\
  \hspace{0.1cm} $e$                                   & 0.0051$\pm$0.0013          \\
  \hspace{0.1cm} $\omega_{\rm 0}$ ($^{\circ}$)         & 129$\pm$16               \\
  \hspace{0.1cm} $K_{1}$ (\kms)                        & 72.23$\pm$0.13             \\
  \hspace{0.1cm} $K_{2}$ (\kms)                        & 77.95$\pm$0.13             \\
  \hspace{0.1cm} $a$ (R$_{\odot}$)\tablenotemark{a}    & 3.7634$\pm$0.0046           \\
  \hspace{0.1cm} $\gamma$ (\kms)                       & $-$118.24$\pm$0.07\tablenotemark{b} \\
  \hspace{0.1cm} $M_{2}/M_{1}$                         & 0.9267$\pm$0.0023           \\
\\
\multicolumn{2}{l}{RMS residuals from the fits}                       \\
  \hspace{0.1cm} Primary (\kms)                        & 1.30                       \\
  \hspace{0.1cm} Secondary (\kms)                      & 1.40                       \\
 \tableline
 \end{tabular}
\tablenotetext{a}{De-projected by adopting the inclination angle from the light}
\tablenotetext{}{curve solutions (see Table~\ref{tab:fit_lc}).}
\tablenotetext{b}{The true uncertainty of $\gamma$ may be larger due to external errors.}
\end{center}
\end{table}

\begin{figure}[t]
\centering
\includegraphics[width=\columnwidth]{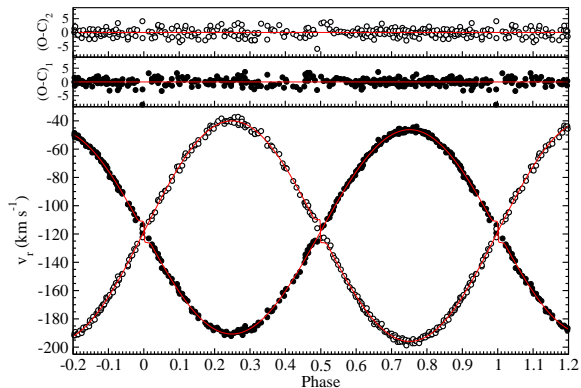}
\caption{Phase-folded radial velocity curves of CM~Dra, with the
primary shown with filled symbols and the secondary with open symbols.
Residuals are shown at the top, and the elements of the spectroscopic
fit are given in Table~\ref{tab:fit_rv}.}
\label{fig:fit_rv}
\end{figure}

Table~\ref{tab:fit_lc} presents the model fits to the different data
sets, with the results from all FCAPT seasons combined
into a single solution. The final column lists our adopted solution in
which we have taken the weighted average of each parameter, with
weights assigned according to the RMS residuals of the fits. The
formal errors reported for the averages are our more conservative
estimates, computed as the quadratic sum of the standard deviation
from the different fits and the internal maximum (statistical) error given by
the WD code for each parameter. The parameters from the fit to the
radial velocity curves are listed in Table~\ref{tab:fit_rv}. The
results for the eccentricity and $\omega$ are consistent with those
derived from the light curves. The fitted light and velocity curves
are shown in Fig.~\ref{fig:fit_lc} and Fig.~\ref{fig:fit_rv}.

The parameters from our light curve fits are generally similar to
those reported by \cite{Lacy1977} (and \citealt{Metcalfe1996}, who
adopted Lacy's photometric results), with the exception of the
relative radius for the secondary, $r_2$. Our value is 2.1\% larger 
than that determined by \cite{Lacy1977}. This discrepancy is significant, 
corresponding to 2 times the combined uncertainities.
One possible explanation is numerical differences in the modeling
techniques: Lacy used the \cite{Russell:52} method, whereas we used
WD. Another is the treatment of the spots: Lacy assumed the spot
modulation to be sinusoidal, whereas we performed a more sophisticated
modeling with WD.  Significant differences in the shape of the
modulation occur near the eclipse phases, as shown in
Fig.~\ref{fig:spot_mod}, which can influence the detailed shape of the
eclipses on which the relative radii depend.  Additionally, Lacy
considered the orbit of CM~Dra to be circular, whereas it is now known
to be slightly eccentric. Because of the impact of $r_2$ on the
absolute dimensions of the binary, we have investigated this
difference by performing a number of light-curve solutions based on
Lacy's data. For this we used the modeling code EBOP \citep[Eclipsing
Binary Orbit Program;][]{Etzel:81, Popper:81}. Under the same
assumptions adopted by Lacy (sinusoidal correction for spots, circular
orbit) we obtain results very close to his for all parameters,
indicating the numerical technique for the modeling is relatively
unimportant.  For a circular orbit but a rectification for spots
computed with WD, as we have done in our own fits, the results differ
somewhat from Lacy's, particularly in $r_2$ but also slightly in the
sum of the relative radii. The largest difference, however, is seen
when abandoning the assumption of a circular orbit. We conclude that
this effect, with some contribution from the treatment of spots, has
introduced subtle biases in the results of \cite{Lacy1977} and
\cite{Metcalfe1996} that are avoided in the present analysis, and gives
us confidence in the accuracy of the absolute properties described
below.

\section{Analysis of the times of minimum}
\label{sec:tmin}
\subsection{Apsidal motion}
\label{subsec:apsidal}

As mentioned in the previous section, CM~Dra has a very small but
significant orbital eccentricity, seen not only in the light curves
but also in the radial-velocity curves. Both General Relativity and
the classical theory of tides predict that a close system such as this
should experience a certain degree of periastron advance. Despite our
attempts described earlier, we were unable to detect a significant
apsidal motion rate ($\dot{\omega}$) in our light curve solutions,
even though those data span nearly 30 years. However, additional
information is available in the form of eclipse timings for both
minima, and we examine these measurements carefully below to
investigate possible changes in the separation between the primary and
secondary eclipses that would be indicative of apsidal motion.

Numerous eclipse timings for CM~Dra have been reported in the
literature using a variety of techniques, beginning with those of
\cite{Lacy1977}. Photoelectric or CCD measurements have greater
precision and are the most useful for our purposes.  Several timings
were obtained in the FCAPT and Sleuth observation campaigns and 
additionally, new timing measurements have been made here with a 
number of telescope facilities, as follows.

A total number ot 20 minima were obtained at the Ond\v{r}ejov observatory 
with the 65-cm reflecting telescope with the Apogee AP-7 CCD camera in 
primary focus. The measurements were done using the Cousins $R$ filter 
with 30~s exposure time. The nearby star GSC~3881.1146 on the same frame 
was selected as a primary comparison. No correction for differential 
extinction was applied because of the proximity of the comparison stars to 
the variable and the resulting negligible differences in airmass. The new 
precise times of minima and their errors were determined by fitting the 
light curves with polynomials.

8 CCD minima were obtained during 2007 and 2008 in the Sloan $r'$ band 
using the 2.0m Liverpool Telescope in La Palma. High quality photometry 
(3--4 mmag per image) was obtained, with typically 100 photometric points 
per event. 63 CCD minima were obtained at the Bradstreet Observatory of 
Eastern University. The equipment consisted of a 41-cm f/10 
Schmidt-Cassegrain reflector coupled to a Santa Barbara Instruments Group 
ST-8 CCD camera binned so as to give a scale of 0.93\arcsec~pixel$^{-1}$. 
All observations were taken through a Cousins $I$ filter. The comparison 
star used was GSC~3881.421 which was always contained within the same 
13$\arcmin$$\times$9$\arcmin$ field.  The exposure times were 25 sec in 
duration, typically resulting in uncertainties of 3 mmag for each data 
point. Finally, a secondary eclipse of CM~Dra was measured with the 1.2-m 
telescope at the F.\ L.\ Whipple Observatory in Arizona using a 
4K$\times$4K CCD camera (KeplerCam), binned to provide a scale of 
0.67\arcsec~pixel$^{-1}$. Observations were made through a Harris $I$ 
filter relative to a set of 30 comparison stars, and exposure times were 
30 sec each. Photometric measurements were performed with IRAF using an 
aperture of 6\arcsec, and have typical uncertainties of 2 mmag. In these 
three latter cases, times of minima were computed by using the 
\citet{Kwee1956} method.

All of these measurements (including those from the literature) have been 
converted to the uniform TT scale, and are presented in 
Table~\ref{tab:minima_data}, which contains a total of 101 primary timings 
and 99 secondary timings. Eclipse timing events coming from different 
sources were weight-averaged. These data span approximately 35 years, 
although there is an unfortunate gap in the coverage of nearly 18 years.

\begin{table*}[t]
\begin{center}
 \caption{Photoelectric and CCD eclipse timings for CM~Dra. The full version of this table is
provided in electronic form.}
 \label{tab:minima_data}
 \begin{tabular}{c c c c l}
 \tableline\tableline

 HJED      & $(O-C)$ (s) & Error (s) & Prim./Sec. & Ref.\\ 
 \tableline
2441855.75476 & $-$25.4 &  30.2 & II & 2 \\
2442555.90592 & $-$35.6 &  30.2 & II & 2 \\
2442557.80955 &    54.7 &  30.2 &  I & 2 \\
2442607.91053 &    18.9 &  30.2 & II & 2 \\
2442888.85928 &    49.9 &  30.2 &  I & 2 \\
2442893.93299 &    62.9 &  30.2 &  I & 2 \\
2442912.95925 &    98.4 &  30.2 &  I & 2 \\
2442966.86433 & $-$30.8 &  30.2 & II & 2 \\
2442994.76890 & $-$31.6 &  30.2 & II & 2 \\
2449494.63438 &    55.1 &   2.8 &  I & 4 \\
  \tableline
 \end{tabular}
\end{center}
\end{table*}
	
In the presence of apsidal motion the times of minimum can be
described following \cite{Gimenez1995} as
\begin{equation}
 \label{eq:O-C_theory}
 \begin{array}{c}
\displaystyle T_{j}= T_{0} + P \left(E + \frac{j-1}{2} \right) + \\
\displaystyle + \left( 2j-3 \right) A_{1}\frac{eP}{2\pi}\cos \omega + \mathcal{O}\left( e^{2} \right),
 \end{array}
\end{equation}
where $j$ indicates a primary or secondary eclipse (1 or 2,
respectively), $E$ is the cycle number, and $A_1$ is a coefficient
dependent on the inclination and eccentricity. The first two terms
represent the linear ephemeris, and the third is the contribution of
the apsidal motion. Given that the eccentricity of CM~Dra is very
small, powers of $e^2$ or higher in these equations have been ignored
since they produce corrections only of the order of 0.2~sec, which are
much below the measurement errors of the timings.
 
For CM~Dra we find that $A_{1} \approx 2$, since the inclination is close
to 90$^\circ$ and the eccentricity is small. Eq.~\ref{eq:O-C_theory}
predicts that the deviation of the times of minimum from a linear
ephemeris has a sinusoidal shape with a semiamplitude of
$\sim$188~sec, and a 180$^{\circ}$ phase difference between the
primary and secondary.  Assuming the rate of periastron advance is
constant, we may write $\omega = \omega_{0} + \dot{\omega} \cdot E$,
where $E$ represents the orbital cycle and $\dot{\omega}$ is the total
apsidal motion of the system. The latter can be determined from fits
of Eq.~\ref{eq:O-C_theory} to each type of timing measurement. In the
approximation of small values of $\dot{\omega} E$,
Eq.~\ref{eq:O-C_theory} can be written as
\begin{equation}
 \label{eq:linear_O-C}
 \begin{array}{c}
\displaystyle T_{j} \simeq T_{0} + P \left(E + \frac{j-1}{2} \right) +\\
\displaystyle + \left(2j-3 \right) A_{1}\frac{eP}{2\pi} \left( \cos \omega_{0} -\sin \omega_{0} \cdot \dot{\omega} E \right).
 \end{array}
\end{equation}
A linear fit to the timings can thus be performed as
\begin{equation}
 \label{eq:linear_fit_O-C}
 \left( O-C  \right)_{j} = \mathcal{T}_{0,j} + \mathcal{P}_{j} E, \\
\end{equation}
where $\mathcal{T}_{0,j}$ can be taken to represent an effective epoch
of reference for both minima including the effect of the eccentricity,
and $\mathcal{P}_{j}$ plays the role of a period for each type of
minimum. The ephemeris for the binary can then be written as
\begin{eqnarray}
 \label{eq:P_Tmin}
 P&=&\frac{\mathcal{P}_{1}+\mathcal{P}_{2}}{2} \\
 T_{0}&=&\frac{\mathcal{T}_{0,1}+\mathcal{T}_{0,2}}{2}-\frac{P}{4}.
\end{eqnarray}
Given values for the orbital elements ($P$, $e$) and $A_{1}$, we
may compute $\omega_{0}$ and $\dot{\omega}$ from the linear fit
parameters $\mathcal{T}_{0,j}$ and $\mathcal{P}_{j}$ as
\begin{eqnarray}
 \label{eq:dif_O-C1}
 \omega_{0} & = & \arccos \left( \frac{2 \pi}{A_{1}eP} \frac{\mathcal{T}_{0,2}-\mathcal{T}_{0,1} - \frac{P}{2}}{2} \right), \\
 \label{eq:dif_O-C2}
 \dot{\omega} & = & \left( \frac{2 \pi}{A_{1}eP \sin \omega_{0}} \frac{\mathcal{P}_{1}-\mathcal{P}_{2}}{2} \right).
\end{eqnarray}
From Eq.~\ref{eq:dif_O-C2} it can be seen that a difference in
the periods for each type of minimum is an indication of the presence
of apsidal motion in the binary.

\begin{figure}[t]
\centering
\includegraphics[width=\columnwidth]{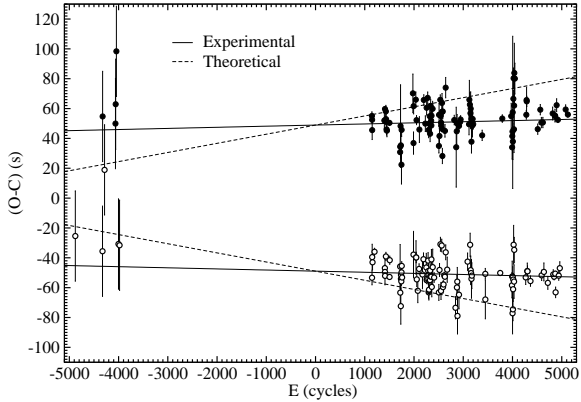}
\caption{Observed minus calculated ($O\!-\!C$) residuals from the
eclipse timings of CM~Dra (filled symbols for the primary, open
symbols for the secondary) with respect to a linear ephemeris. The
linear fits to apsidal motion (solid line) and the
theoretically predicted apsidal motion (dashed line) are shown.}
\label{fig:tmin_O-C}
\end{figure}

Fig.~\ref{fig:tmin_O-C} shows the $O\!-\!C$ values for the primary and 
secondary minima of CM~Dra as a function of the cycle number. There would 
seem to be a linear trend although the scatter of the measurements is 
fairly large.  This scatter may be due in part to random errors, but there 
could also be biases arising from the presence of spots on the surface of 
the components. As a test, we simulated light curves for CM~Dra with the 
different spot configurations given in Table~\ref{tab:spots}, and we found 
that the presence of spots can indeed skew eclipse timing determinations 
by up to $\sim 15$~sec. Similar results were found in a study by 
\citet{Hargis2000}. Because of this effect, observational errors in the 
timings could be significantly underestimated. We therefore performed 
linear fits of the times of eclipse with the internal errors doubled, in 
order to preserve the relative weights between the measures and obtain a 
reduced $\chi^{2}$ value closer to unity. This yielded more realistic 
uncertainties for the parameters of the fit. The results are shown in 
Table~\ref{tab:fit_O-C}, and indicate an apsidal motion of $\dot{\omega} = 
(2.3 \pm 1.4)\cdot 10^{-4}$~deg~cycle$^{-1}$, i.e., a detection with 
1.6$\sigma$ significance. Tests in which the internal errors were 
augmented by adding 15~sec in quadrature (to account for the potential 
effects of spots) instead of doubling them gave the same results, within 
the errors.

\begin{table}[t]
 \begin{center}
 \caption{Results of the linear fits to the eclipse timings for apsidal motion.}
 \label{tab:fit_O-C}
 \begin{tabular}{l c}
 \tableline\tableline
 Properties          & Weighted fit                           \\
 \hline                                                     
  $\mathcal{T}_{0,1}$ (s)               & 48042.32778$\pm$0.00002        \\
  $\mathcal{P}_{1}$ (s cycle$^{-1}$)  & 1.2683899936$\pm$0.0000000064   \\
  $\chi^{2}_{1}$                      & 1.303                 \\
  $ $ \\
  $\mathcal{T}_{0,2}$ (s)               &48042.96084$\pm$0.00002        \\
  $\mathcal{P}_{2}$ (s cycle$^{-1}$)  & 1.2683899765$\pm$0.0000000069 \\
  $\chi^{2}_{2}$                      & 0.920               \\
  $ $ \\
  $P$ (days)                  & 1.268389985$\pm$0.000000005         \\
  $T_{0}$ (HJED)              & 48042.327214$\pm$0.000014         \\
  $ $ \\
  $\omega_{0}$ (deg)                  & 104.9$\pm$3.7         \\
  $\dot{\omega}$ (deg cycle$^{-1}$) & (2.3$\pm$1.4) 10$^{-4}$ \\
  $U$ (years)                         & 5400$\pm$3200         \\
 \tableline
 \end{tabular}
\end{center}
\end{table}

\subsection{Third body effects on the eclipse timings}
\label{sec:third}

The analysis of times of minimum can also reveal the presence of third 
bodies in eclipsing systems through the time-delay effect caused by the 
orbit of the binary around the barycenter of the system. This produces a 
sinusoidal modulation of the $(O\!-\!C)$ values from the timings. 
\citet{Deeg2008} have recently reported the possible presence of a third 
body around CM~Dra based on a parabolic fit to their sample of $(O\!-\!C)$ 
values. We find, however, that using our own timings a parabolic fit is 
essentially indistinguishable from a linear fit to the measurements. Thus, 
any third body must have a period longer than roughly twice the span of 
the measurements, or $\sim$60 years, or must induce a light-time effect 
below $\sim 15$~s which would be undistinguishable from the dispersion of 
the data due to spot effects.

\begin{figure}[t]
\centering
\includegraphics[width=\columnwidth]{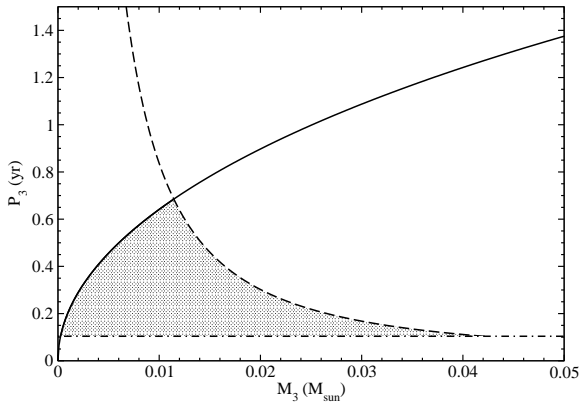}
\caption{Allowed region (shaded area) in a mass vs. orbital period 
diagram for a third
body in the CM~Dra system according to observational constraints: 
eccentricity modulation (solid line), light-time effect (dashed line)
and stability criteria (dot-dashed line).}
\label{fig:limit_M3}
\end{figure}

Another indication of the possible presence of a third body is the
small eccentricity of the close binary orbit of CM~Dra. Systems with
periods as short as that of CM~Dra are usually assumed
to be tidally circularized early on \citep{Mazeh2008}, possibly even
during the pre-main sequence phase. To explain the present non-zero
eccentricity one may invoke the presence of a perturbing component in
a more distant orbit.  Such a configuration can produce secular
variations of the orbital parameters of the inner orbit, such as an
eccentricity modulation with a typical period $U_{mod}$ given by
\begin{equation}
\label{eq:third_modulation}
U_{mod}\simeq P_{1,2} \left( \frac{a_{3}}{a_{1,2}} \right)^{3}\frac{M_{1}+M_{2}}{M_{3}},
\end{equation}
where $P_{1,2}$ and $a_{1,2}$ are the period and semimajor axis of the 
inner orbit of CM~Dra, and $a_{3}$ and $M_{3}$ are the semimajor axis of 
the third body around the center of mass of the triple system and the mass 
of the third body, respectively. A third body is actually known in the 
CM~Dra system (the common proper motion white dwarf companion).  Adopting 
a mass for the white dwarf of 0.63~M$_{\odot}$ from \cite{Bergeron2001}, 
along with an angular separation from CM~Dra of about 26\arcsec\ 
(corresponding to $\sim$380~AU at the distance of CM~Dra), the modulation 
period on CM~Dra would be roughly 2~Gyr. However, the effect of such a 
long-period eccentricity pumping would be averaged out over many apsidal 
motion cycles, and therefore the orbit would remain circular. One may 
assume that eccentricity pumping by some other body in the system will 
only be effective if $U_{mod} \lesssim 5400$ years, which is the period of 
the apsidal motion found for CM~Dra. This provides a constraint on the 
properties of this putative body, if it is to explain the measured 
eccentricity. Fig.~\ref{fig:limit_M3} represents the allowed region (mass 
vs. period) of the companion by accounting for the non-detection of 
light-time effect above $15$ sec and the eccentricity pumping. Also, we 
consider that $P_{3}/P_{\rm CM~Dra} \gtrsim 30$ for stability criteria of 
hierarchical triple systems. As can be seen, a massive planet or light 
brown dwarf with an orbital period of 50--200 days would fulfill all 
constraints.

\section{Absolute properties of the components of CM~Dra}
\label{sec:properties}

Based on the fits to the light curves and the radial velocities, the
absolute physical properties of the components of CM~Dra including the
masses and radii can be derived independently of distance or flux
calibrations. We report these values in Table~\ref{tab:physical}. With
the measured radii, we find that the predicted rotational velocities
of the primary and secondary, assuming synchronous rotation, are
$10.22\pm0.08$ \kms\ and $9.67\pm0.07$ \kms, respectively. These
values are in good agreement with the $v \sin i$ measurements from our
spectra (\S\,\ref{subsec:vc}).

The effective temperatures of the components are not directly
accessible from the light curve analysis, which yields only their
ratio as measured by the relative depths of the eclipses. In
\S\,\ref{sec:lc_rv} we adopted a value for the primary $T_{\rm eff}$
from an external source \citep{Viti1997, Viti2002}, based on a
modeling of the spectrum of CM~Dra. It is possible, however, to
determine the individual temperatures in another way, using
information from the light curves along with a combined near-IR
magnitude for the system and its trigonometric parallax. Here we have
used the 2MASS magnitude $K_s = 7.796 \pm 0.021$, subsequently
converted to the Johnson system following \cite{Carpenter:01}, and the
parallax $\pi = 69.2 \pm 2.5$ mas from \cite{vanAltena1995}.  We chose
to rely on a near-IR magnitude because the corresponding bolometric
corrections are less dependent on $T_{\rm eff}$ and chemical
composition.  We began by adopting the value of $T_{\rm eff} = 3100$~K
as a starting point for the primary, from which the secondary $T_{\rm
eff}$ follows from the measured temperature ratio. Bolometric
corrections for each star were taken from \cite{Bessell1998} as a
function of temperature, and averaged since they are virtually
identical. The total luminosity was then computed. The ratio of the
luminosities can be calculated from the temperature ratio and radius
ratio, both of which are measured directly and accurately from the
light curves:
\begin{equation}
 \label{eq:ratio_L}
 \frac{L_{2}}{L_{1}}=\left( \frac{r_{2}}{r_{1}}  \right)^{2} \left( \frac{T_{\rm eff, 2}}{T_{\rm eff,1}}  \right)^{4}=0.880\pm0.022.
\end{equation}
Individual bolometric luminosities are thus easily derived, and since
the absolute radii are also known, the individual temperatures can be
obtained. This process was iterated until the corrections to the
temperatures were below 1~K. The result is independent of the starting
point for the primary temperature. The mean bolometric correction
resulting from the calculation is $BC_{\rm K} = 2.66 \pm 0.05$, and
the total luminosity is $0.0104 \pm 0.0009$~L$_{\odot}$. The individual
temperatures and luminosities are listed in Table~\ref{tab:physical},
in which the uncertainties include all measurement errors as well as
an assumed uncertainty of 0.05 mag for the bolometric corrections, but
exclude systematics that are difficult to quantify. The $T_{\rm eff}$
values, which have a mean of 3125~K, agree very well with the estimate
of \cite{Viti1997, Viti2002}.

\begin{table}[t]
 \begin{center}
 \caption{Absolute physical properties of CM~Dra.}
 \label{tab:physical}
 \begin{tabular}{l c c}
 \tableline\tableline
 Properties          & Component 1        & Component 2       \\
 \hline                                                     
  $M$ (M$_{\odot}$)  & 0.2310$\pm$0.0009  & 0.2141$\pm$0.0010 \\
  $R$ (R$_{\odot}$)  & 0.2534$\pm$0.0019  & 0.2396$\pm$0.0015 \\
  $\log g$ (cgs)     & 4.994$\pm$0.007    & 5.009$\pm$0.006   \\
  $T_{\rm eff}$ (K)  & 3130$\pm$70        & 3120$\pm$70       \\
  log($L/L_{\odot}$) & $-$2.258$\pm$0.038 & $-$2.313$\pm$0.056\\
  Age (Gyr)          & \multicolumn{2}{c}{$4.1\pm0.8$ (Main Sequence)}\\
  $[M/H]$            & \multicolumn{2}{c}{$-1 < [M/H] < -0.6$} \\
 \tableline
 \end{tabular}
\tablenotetext{}{$M_{\rm Bol \odot}$=4.74 is used to compute luminosities \citep{Bessell1998}.}
\end{center}
\end{table}

As a check, we used the above temperatures and our light ratios in the
$R$ and $I$ bands from the light curves to predict the light ratio in
$V$, appropriately scaling the NextGen models \citep{Hauschildt1999}. 
The result
is $L_2/L_1 = 0.86 \pm 0.15$, which is consistent with the value
determined spectroscopically (\S\,\ref{subsec:vc}), within the errors.
The mean temperature of the system may also be estimated from
available color indices for CM~Dra, and the recent color/temperature
calibration for M dwarfs by \cite{Casagrande:08}. We used the $VR\,I$
magnitudes of \cite{Lacy1977} and the $JHK_s$ magnitudes from 2MASS to
construct twelve different color indices \citep[after conversion of
the $R\,I$ magnitudes from the Johnson system to the Cousins system
following][] {Leggett:92}, which are of course not independent of each
other although they do serve to gain a better idea of the scatter
among the various calibrations. We obtain a weighted average
temperature of $3050 \pm 50$~K, which is only slightly lower than the
estimates above, but has the virtue of being completely independent
of the parallax and the light curve parameters.

The age, along with mass and chemical composition, is an indicator of
the evolutionary status of a star. When known, it becomes a powerful
constraint that can be used in the model comparisons. For CM~Dra we
may obtain a rough estimate of its age by considering the properties
of its white dwarf companion. According to \citet{Bergeron2001}, the
cooling age of the white dwarf is $2.84\pm0.37$~Gyr. Given its
estimated mass ($\sim$0.63~M$_{\odot}$) and the initial-final mass
relationship of \citet{Catalan2008}, the mass of the main sequence
progenitor is estimated to be $2.1\pm0.4$~M$_{\odot}$. For stars of such
mass, stellar evolution models predict a lifetime of about 1.3~Gyr
\citep{Girardi2000}. We therefore
infer an approximate age for CM~Dra of 4.1~Gyr with a 20\% uncertainty
level coming from uncertainties in the mass of the white dwarf progenitor
and its metallicity. This total age indicates that CM Dra is
well on the main sequence.

The chemical composition of CM~Dra has been notoriously difficult to
determine, which is unfortunate for such an important system.  It has
usually been considered to be metal-poor, although this is based
mostly on circumstantial evidence (i.e., its large space motion).
Attempts to determine the metallicity by various means have often
produced inconsistent results. \cite{Gizis:97} and \cite{Leggett:98}
concluded the composition is near solar, while \citet{Viti1997,
Viti2002} found a metal-poor composition ($-1.0 < [M/H] < -0.6$) by
performing fits to the spectral energy distribution and several
diagnostic spectral features using stellar atmosphere models. However,
some systematic differences between the estimates from optical and
near-IR spectra in the latter studies are disconcerting and cast some
doubts on the results. Our own checks using the same spectroscopic
material and the most recent version of the NextGen models did not
yield an improvement in the results. The various metallicity
indicators still show disagreements, and would seem to indicate
shortcomings in the model atmosphere calculations. Thus, the
metallicity of CM~Dra remains poorly determined.

Kinematics of the common proper motion group of CM~Dra could provide 
further insight on its age and metallicity. The space velocity components 
of the system are $U=-106.8$~\kms, $V=-119.8$~\kms and $W=-35.1$~\kms. 
These values indicate that the system probably does not belong to the thin 
disk population. No clear correlation between kinematics and metallicity 
or age has been found for stars on the solar neighborhood 
\citep{Nordstrom2004}, and certainly no claims can be made on an 
individual star basis. We must conclude that the kinematics of CM~Dra do 
not seem to stand in contradiction with an age of about $4$ Gyr, neither 
help to discern between a solar or moderately sub-solar metallicity. 
Another interesting trait of CM Dra is the fact that it has remained 
weakly bound to its reltively distant companion for a long time. The 
binding energy of the system is over four orders of magnitude smaller than 
its kinetic energy with respect to the local standard of rest. Whatever 
perturbations the system has suffered during its life, they must come from 
a smooth potential or else the pair would have been broken. This should 
provide interesting constraints to the mechanisms of star acceleration in 
the Galaxy.

\section{Comparison with theoretical models}
\label{sec:models}

\begin{figure*}[t]
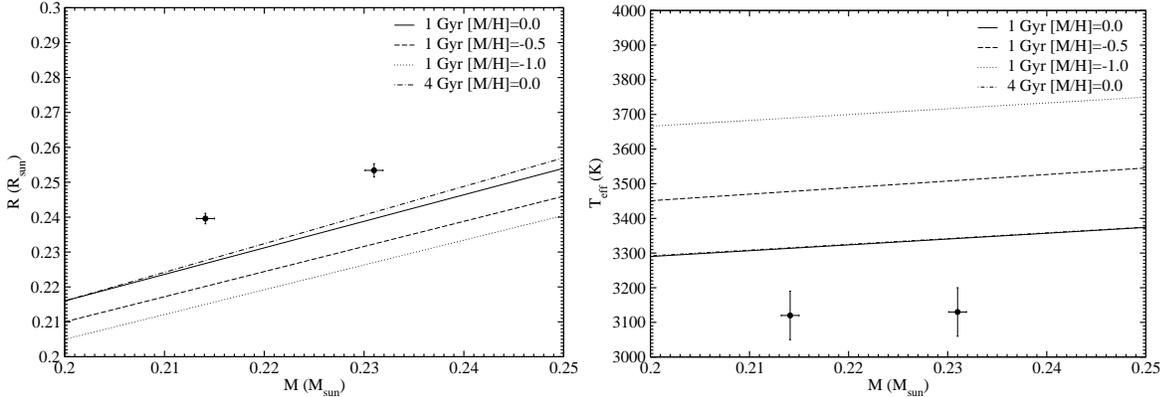

\centering
\includegraphics[width=\columnwidth]{f7a.eps}
\includegraphics[width=\columnwidth]{f7b.eps}
\caption{Comparison of the physical properties for CM~Dra with 
models of \citet{Baraffe1998} in the $M$-$R$ diagram (left) and the
$M$-$T_{\rm eff}$ diagram (right). Models for different metallicities
and ages are shown, as labeled.}
\label{fig:models}
\end{figure*}

Our mass, radius, and temperature determinations for CM~Dra are
compared in Fig.~\ref{fig:models} with the predictions of stellar
evolution models from \citet{Baraffe1998}. The measurements support a
trend found previously for other low-mass EBs, in the sense that the
observed radii for both components are larger than predicted by
theory, in this case by $\sim$4.7\% and $\sim$5.0\% for the primary
and secondary, respectively. The effective temperatures are cooler
than the models indicate, by $\sim$6.8\% and $\sim$6.3\%. We note also
that while these are significant \emph{offsets} (compared to the
errors), the \emph{slope} of the models appears substantially
correct. These deviations refer strictly to the comparison with
solar-metallicity models, and would be even larger if a lower
metallicity were assumed. For example, the offsets would increase to
$\sim$10\% in the radii for models with $[M/H]=-1$.

Magnetic activity on the components of similar low-mass EB systems has
often been proposed as an explanation for these discrepancies between
models and observations (see \S\,\ref{sec:introduction}) . The 
activity in these typically short-period
binaries is associated with the very rapid rotation resulting from
tidal synchronization with the orbital motion. One manifestation of
this activity is the presence of surface features (spots) that tend to
block a fraction of the outgoing radiation. The star adjusts by
increasing its size in order to conserve flux, and at the same time
the effective temperature becomes lower than in a spot-free star.
Recent work has shown that the same hypothesis appears to explain the
differences observed between active and inactive single stars
\citep{Morales2008}. Theoretical efforts have had some success in
reproducing the observations for sub-solar mass binary systems by
accounting for stellar activity in the models, at least to first order
\citep[e.g.,][]{DAntona:00, Mullan:01, Chabrier2007}. In the most
recent of these studies the authors examined the effects of activity
in reducing the convection efficiency as well as in obstructing
radiation due to the presence of dark surface features. The first of
these effects is equivalent to a reduction in the mixing length
parameter ($\alpha_{\rm ML}$), whereas the second can be parametrized
in terms of the fractional spot coverage. The results show that for
stars in the fully convective regime ($M \lesssim 0.35$~M$_{\sun}$)
the effect of a reduction in $\alpha_{\rm ML}$ is minimal, while the
presence of spots has a significant effect, and accounting for this
can in fact reproduce the properties of CM~Dra with a spot fraction of
about 30\%. For somewhat more massive stars theory predicts that both a reduced
$\alpha_{\rm ML}$ and spot coverage lead to similar effects on the
global properties. Although further observational and theoretical work
is needed, these predictions appear at least qualitatively consistent
with the findings of previous EB studies that suggest the radius
discrepancies with the models are roughly 5\% for stars with $M <
0.35$~M$_{\sun}$, and about 10\% for higher mass stars with convective
envelopes. The implication is that in the former case the deviations 
are due only to spots, whereas for stars with radiative cores both spots 
and the reduction in the convection efficiency are important. A more 
detailed study of the relationship between activity and the radius
discrepancies is underway by a subset of the present authors,
including consistency checks with all observational constraints for
late-type stars in binaries. This work will be presented in a
forthcoming paper, placing CM~Dra in context with the rest of the
low-mass EBs.

The value of the apsidal motion found in \S\,\ref{sec:tmin}
provides a different type of test of internal structure models since
the rate of classical precession induced by tidal effects depends on
the density profile of the stars. Following the prescriptions in 
\citep{Kopal1978}, tidal apsidal motion is given by:
\begin{equation}
 \label{eq:dwdt_tidal}
 \dot{\omega_{tidal}}=360^{\circ} \left( c_{2,1}k_{2,1}+c_{2,2}k_{2,2} \right),
\end{equation}
where $c_{2,i}$ are coefficients that depend on the properties of 
each component, and $k_{2,i}$ are the internal structure constants dependent
on the density profile. Using the internal profile models of 
\citet{Baraffe1998}, we derive $\log k_{2,1}=-0.95$ and 
$\log k_{2,2}=-0.96$ for the primary and secondary components, respectively.
These values yield  $\dot{\omega_{tidal}}=\left( 1.64\pm0.04 \right)\cdot 
10^{-3}$~deg~cycle$^{-1}$.

However, several more phenomena can contribute to the magnitude of the 
total apsidal motion, aside from the classical effects of tidal 
interaction. One is relativistic precession, which depends essentially on 
the masses and orbital period of the binary system. Following the formula 
given by \citet{Gimenez1985}, and based on the properties derived here for 
CM~Dra, the predicted effect is $\dot{\omega}_{\rm rel} = (2.711 \pm 
0.005)\cdot 10^{-4}$~deg~cycle$^{-1}$. Combining the tidal and 
relativistic contributions, we obtain a theoretical value of the total 
apsidal motion of $\dot{\omega_{theo}}=\left( 1.91\pm0.04 \right)\cdot 
10^{-3}$~deg~cycle$^{-1}$. This value is incompatible with the observed 
$\dot{\omega_{obs}}=\left( 2.3\pm1.4 \right)\cdot 
10^{-4}$~deg~cycle$^{-1}$. The discrepancy is significant, at the 
12-$\sigma$ level.

A third body may also alter the apsidal motion rate of the binary. 
Interaction with the distant white dwarf companion needs in principle some 
consideration. The typical period of the apsidal motion resulting from a 
third-body perturbation is given by Eq.~\ref{eq:third_modulation}. As 
described earlier, for the white dwarf CM Dra companion this modulation 
period is about 2~Gyr, thus the contribution of the white dwarf to the 
apsidal motion is completely negligible compared to the relativistic 
contribution, which has a much shorter period of $\sim$4600~yr. There have 
been some claims of detections of low-mass companions orbiting CM~Dra from 
analyses of the eclipse timings \citep[e.g.,]{Deeg2000, Deeg2008}, but the 
evidence so far does not seem compelling. As we discuss in \S 
\ref{sec:third} a third body may help to explain the small but 
signifincant eccentricity and remain undetected to via the light-time 
effect. Given the possible range in mass and orbital period of such 
putative substellar object, we estimate that its contribution to the 
apsidal motion of CM~Dra may be sufficient (depending on the relative 
orbital inclination) to explain the observed differences. Note that 
scenario has been advocated to resolve the discrepancy found in the 
eclipsing binary DI Her \citep{Guinan1985}.

CM~Dra has often been regarded in the past as a favorable system for
inferring the primordial helium abundance, assuming that it is a
Population~II star. \citet{Paczynski1984} described a method using
polytropic stellar models that was followed by \citet{Metcalfe1996},
who obtained bulk helium abundances of about 0.3 for both stars. These
values are significantly higher than estimates using other
methods. Another analysis by \citet{Chabrier1995} led to a much lower
value of 0.25, through a comparison between their models and the
physical properties of the CM~Dra components reported by
\citet{Lacy1977}. Those authors indicated, however, that for masses as
low as those of CM~Dra the models depend only weakly on the helium
abundance.  The present work shows that standard models are as yet
unable to reproduce the observed values of $R$ and $T_{\rm eff}$ for
the components of this binary at their measured masses. It would seem,
therefore, that a determination of the helium abundance by comparison
with model predictions (or simpler polytropes) is not particularly
meaningful at the moment, given that the significant effects of
activity on the global properties of these stars are not yet properly
accounted for.

\section{Conclusions}
\label{sec:conclusions}

Prompted by significant improvements in eclipsing binary analysis
methods since the most recent major photometric study by
\cite{Lacy1977}, and aided by new photometric observations
gathered here to complement existing data, as well as improvements in
the spectroscopy, we have conducted a thorough reanalysis of the
classical low-mass double-lined eclipsing binary CM~Dra. The goal
has been to provide the best possible determinations of the physical
properties to enable stringent tests of stellar theory. Our results
for the masses and radii of the stars, which we estimate to be about
4.1~Gyr old, are $M_{1}=0.2310\pm0.0009$~M$_{\odot}$,
$M_{2}=0.2141\pm0.0010$~M$_{\odot}$,
$R_{1}=0.2534\pm0.0019$~R$_{\odot}$, and
$R_{2}=0.2396\pm0.0015$~R$_{\odot}$, with formal relative
uncertainties of only $\sim$0.5\%. A special effort has been made in
this study to investigate possible sources of systematic error in
these quantities and to assess their importance. We have performed a
number of tests during the light-curve analysis, the spectroscopic
analysis, and the determination of effective temperatures. The
resulting uncertainties of these physical properties are thus believed
to be realistic, and to offer the best opportunity for carrying out
meaningful tests of models of stellar evolution and stellar structure
for fully convective main-sequence stars.

We find that the radii and temperatures of CM~Dra show the same sort
of discrepancy with model predictions as found previously for other
low-mass EBs, which are at the level of $\sim$5--7\% in this
particular case.  Mounting evidence indicates that such differences
can be ascribed to magnetic activity effects. Further research is
underway to estimate the corrections needed in the models in order to
reproduce the observations of low-mass EBs, given the prescriptions
proposed by \citet{Chabrier2007}.

Measurements of the times of minimum for CM~Dra clearly show the presence 
of apsidal motion in the system. However, its value is still poorly 
determined on account of observational errors, other errors due to 
distortions caused by spots, and the limited time coverage of the data. 
There also seems to exist a discrepancy bewteen the observational value 
and that derived from General Relativity and tidal theory. A third body in 
the system, which may be responsible for the non-zero eccentricity, could 
provide an explanation to the observed difference. Further measurements 
over the coming years will greatly help to constrain the precession of the 
line of apsides and separate the two effects more clearly, in addition to 
providing a better basis for investigating the possible presence of a 
third body in the system.

\acknowledgements

We are grateful to J.\ Kreiner for help with the compilation of eclipse 
timings. JCM, IR\ and CJ\ acknowledge support from the Spanish Ministerio de 
Educaci\'on y Ciencia via grants AYA2006-15623-C02-01 and 
AYA2006-15623-C02-02. GT\ acknowledges partial support for this work from 
NSF grant AST-0708229 and NASA's MASSIF SIM Key Project (BLF57-04). EFG 
acknowledges support for the FCAPT research under NSF/RUI grant 
AST05-07536. DC\ and FTOD\ acknowledge support for the Sleuth Observatory 
work by the National Aeronautics and Space Administration under grant 
NNG05GJ29G issued through the Origins of Solar Systems Program. FTOD\
also acknowledges partial support for this work provided through the NASA 
Postdoctoral Program at the Goddard Space Flight Center, administered by 
Oak Ridge Associated Universities through a contract with NASA. The 
investigation of MW\ was supported by the Grant Agency of the Czech 
Republic under grant No.~205/06/0217.

\end{document}